\documentclass[sigconf]{acmart} 
\copyrightyear{2019}
\acmYear{2019}
\setcopyright{iw3c2w3}
\acmConference[WWW '19]{Proceedings of the 2019 World Wide Web Conference}{May 13--17, 2019}{San Francisco, CA, USA}
\acmBooktitle{Proceedings of the 2019 World Wide Web Conference (WWW '19), May 13--17, 2019, San Francisco, CA, USA}
\acmPrice{}
\acmDOI{10.1145/3308558.3313488}
\acmISBN{978-1-4503-6674-8/19/05}

\def\BibTeX{{\rm B\kern-.05em{\sc i\kern-.025em b}\kern-.08emT\kern-.1667em\lower.7ex\hbox{E}\kern-.125emX}}
\usepackage{amssymb}
\usepackage{algorithm}
\usepackage{algorithmic}
\usepackage{bibentry}
\usepackage{amsfonts}
\usepackage{subfigure}
\usepackage{multirow}
\usepackage{amsmath}
\usepackage{graphicx}
\usepackage{epsfig}
\usepackage{color}
\usepackage{epstopdf}
\usepackage{rotating}
\usepackage{caption}
\usepackage{multicol}
\usepackage{amssymb}
\usepackage{graphicx}
\usepackage{bbding}
\usepackage{algorithm}
\usepackage{algorithmic}
\usepackage{balance}

\begin{document}

\title{Graph Neural Networks for Social Recommendation}

\author{Wenqi Fan}
\affiliation{%
  \institution{Department of Computer Science}
  \state{City University of Hong Kong}
}
\email{wenqifan03@gmail.com}

\author{Yao Ma}
\affiliation{%
  \institution{Data Science and Engineering Lab}
  \state{Michigan State University}
}
\email{mayao4@msu.edu}

\author{Qing Li}
\affiliation{%
  \institution{Department of Computing}
  \state{The Hong Kong Polytechnic University}
}
\email{csqli@comp.polyu.edu.hk}

\author{Yuan He}
\affiliation{%
  \institution{JD.com}
}
\email{heyuan6@jd.com}

\author{Eric Zhao}
\affiliation{%
  \institution{JD.com}
}
\email{ericzhao@jd.com}

\author{Jiliang Tang}
\affiliation{%
  \institution{Data Science and Engineering Lab}
  \state{Michigan State University}
}
\email{tangjili@msu.edu}

\author{Dawei Yin}
\affiliation{%
  \institution{JD.com}
}
\email{yindawei@acm.org	}

\begin{abstract}
In recent years, Graph Neural Networks (GNNs), which can naturally integrate node information and topological structure, have been demonstrated to be powerful in learning on graph data. These advantages of GNNs provide great potential to advance social recommendation since data in social recommender systems can be represented as user-user social graph and user-item graph; and learning latent factors of users and items is the key. However, building social recommender systems based on GNNs faces challenges. For example, the user-item graph encodes both interactions and their associated opinions; social relations have heterogeneous strengths; users involve in two graphs (e.g., the user-user social graph and the user-item graph). To address the three aforementioned challenges simultaneously, in this paper, we present a novel graph neural network framework \textbf{(GraphRec)} for social recommendations. In particular, we provide a principled approach to jointly capture interactions and opinions in the user-item graph and propose the framework GraphRec, which coherently models two graphs and heterogeneous strengths. Extensive experiments on two real-world datasets demonstrate the effectiveness of the proposed framework GraphRec. Our code is available at \url{https://github.com/wenqifan03/GraphRec-WWW19}
\end{abstract}


\maketitle

\section{Introduction}

The exploitation of social relations for recommender systems has attracted increasing attention in recent years~\cite{ma2011recommender, tang2013exploiting, tang2016recommendation}. These social recommender systems have been developed based on the phenomenon that users usually acquire and disseminate information through those around them, such as classmates, friends, or colleagues, implying that the underlying social relations of users can play a significant role in helping them filter information~\cite{resnick1997recommender}. Hence, social relations have been proven to be helpful in boosting the recommendation performance~\cite{DeepSoR2018,tang2013social}.

\begin{figure}[htbp]
\centering
{\includegraphics[width=0.90\linewidth]{{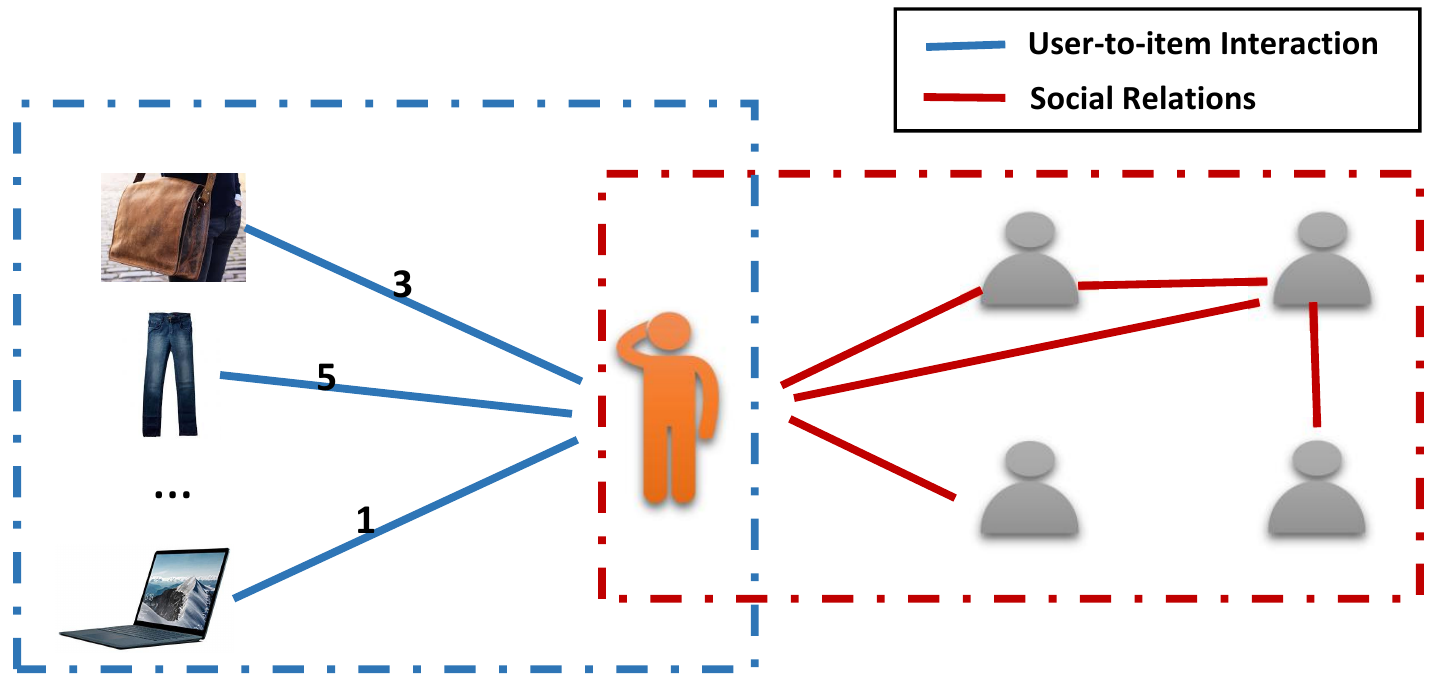}}}
\caption{Graph Data in Social Recommendation. It contains two graphs including the user-item graph (left part) and the user-user social graph (right part). Note that the number on the edges of the user-item graph denotes the opinions (or rating score) of users on the items via the interactions.} \label{fig:Introduction_SocialAggregation}
\end{figure}

Recent years have witnessed great developments in deep neural network techniques for graph data~\cite{kipf2017semi}. These deep neural network architectures are known as Graph Neural Networks (GNNs)~\cite{hamilton2017inductive, defferrard2016convolutional, ma2018multi}, which have been proposed to learn meaningful representations for graph data. Their main idea is how to iteratively aggregate feature information from local graph neighborhoods using neural networks. Meanwhile, node information can be propagated through a graph after transformation and aggregation. Hence, GNNs naturally integrate the node information as well as the topological structure and have been demonstrated to be powerful in representation learning~\cite{kipf2017semi, defferrard2016convolutional,derr2018signed}. On the other hand, data in social recommendation can be represented as graph data with two graphs. As demonstrated in Figure~\ref{fig:Introduction_SocialAggregation}, these two graphs include a social graph denoting the relationships between users, and a user-item graph denoting interactions between users and items. Users are simultaneously involved in both graphs, who can bridge them. Moreover, the natural way of social recommendation is to incorporate the social network information into user and item latent factors learning~\cite{yang2017social}. Learning representations of items and users is the key to build social recommender systems. Thus, given their advantages, GNNs provide unprecedented opportunities to advance social recommendation.

Meanwhile, building social recommender systems based on GNNs faces challenges.  The social graph and the user-item graph in a social recommender system provide information about users from different perspectives. It is important to aggregate information from both graphs to learn better user representations. Thus, the first challenge is how to inherently combine these two graphs. Moreover, the user-item graph not only contains interactions between users and items but also includes users' opinions on items. For example, as shown in Figure~\ref{fig:Introduction_SocialAggregation}, the user interacts with the items of ``trousers" and ``laptop"; and the user likes
``trousers" while disliking ``laptop". Therefore, the second challenge is how to capture interactions and opinions between users and items jointly. In addition, the low cost of link formation in online worlds can result in networks with varied tie strengths (e.g., strong and weak ties are mixed together) ~\cite{xiang2010modeling}. Users are likely to share more similar tastes with strong ties than weak ties. Considering social relations equally could lead to degradation in recommendation performance. Hence, the third challenge is how to distinguish social relations with heterogeneous strengths.


In this paper, we aim to build social recommender systems based on graph neural networks. Specially, we propose a novel graph neural network \textbf{GraphRec} for social recommendations, which can address three aforementioned challenges simultaneously. Our major contributions are summarized as follows:
\begin{itemize}
\item We propose a novel graph neural network GraphRec, which can model graph data in social recommendations coherently;
\item We provide a principled approach to jointly capture interactions and opinions in the user-item graph;
\item We introduce a method to consider heterogeneous strengths of social relations mathematically; and
\item We demonstrate the effectiveness of the proposed framework on various real-world datasets.
\end{itemize}

The remainder of this paper is organized as follows. We introduce the proposed framework in Section~\ref{sec:methodlogy}. In Section~\ref{sec:Experiments}, we conduct experiments on two real-world datasets to illustrate the effectiveness of the proposed method. In Section~\ref{sec:relatedwork}, we review work related to our framework. Finally, we conclude our work with future directions in Section~\ref{sec:conclusion}. 
\section{The Proposed Framework}
\label{sec:methodlogy}
In this section, we will first introduce the definitions and notations used in this paper, next give an overview about the proposed framework, then detail each model component and finally discuss how to learn the model parameters.

\begin{figure*}[htbp]
\centering
{\includegraphics[width=0.99\linewidth]{{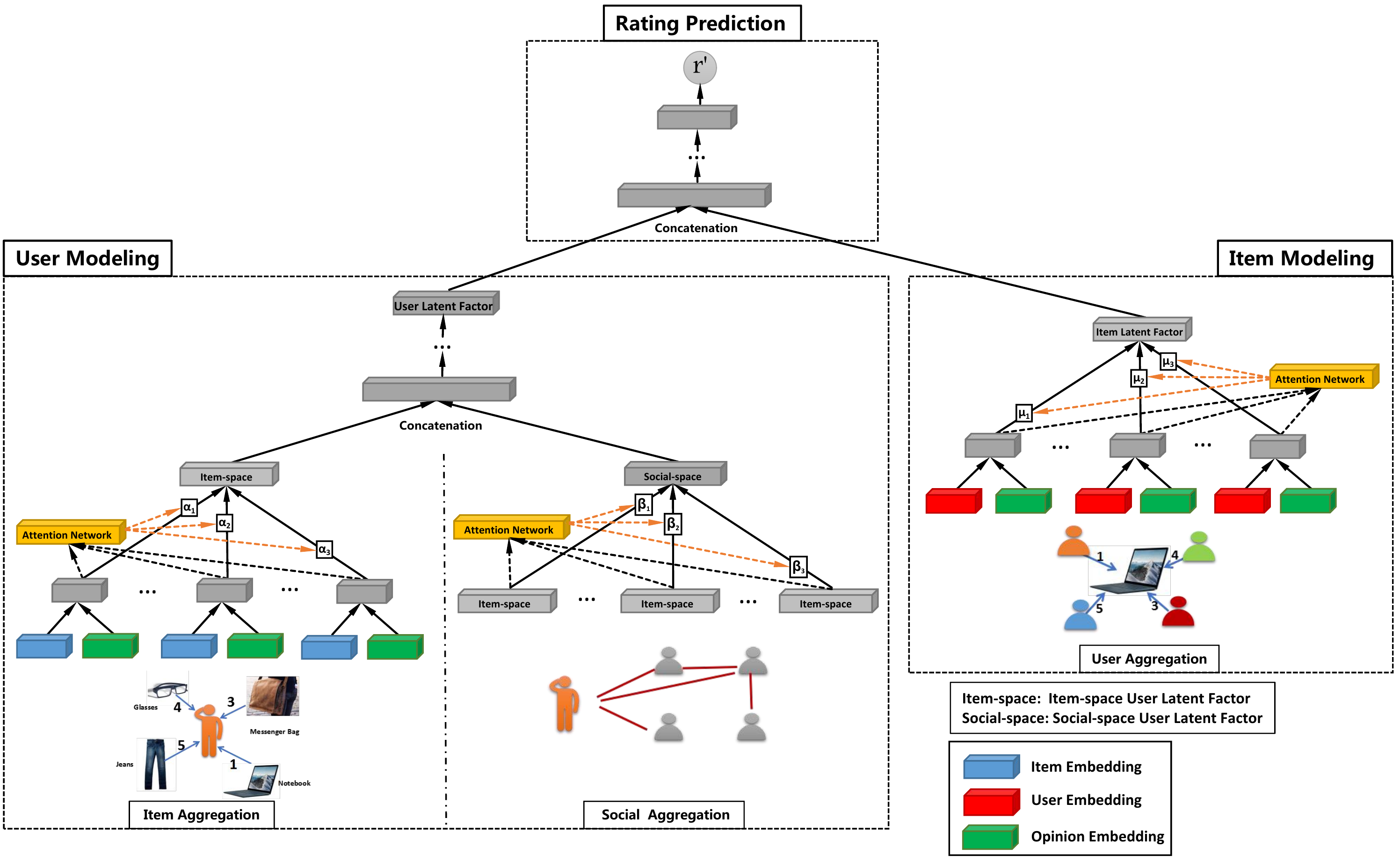}}}
\caption{The overall architecture of the proposed model. It contains three major components: user modeling, item modeling, and  rating prediction.}\label{fig:GraphRec}
\end{figure*}

\begin{table}[htbp]
\centering
\caption{Notation}
\label{tab:notation}
\begin{tabular}{c|c}
Symbols & Definitions and Descriptions \\ \hline
$r_{ij}$        & The rating value of item $v_j$ by user $u_i$           \\ \hline
$\mathbf{q}_j$        & The embedding of item $v_j$                             \\ \hline
$\mathbf{p}_i$        & The embedding of user $u_i$                             \\ \hline
$\mathbf{e}_r$        & \begin{tabular}[c]{@{}c@{}}The opinion embedding for the rating level $r$,\\  such as 5-star rating, $r \in \left \{ 1,2,3,4,5 \right \}$ \end{tabular}    \\ \hline
$d$ & The length of embedding vector \\ \hline
$C(i)$       &  The set of items which user $u_i$ interacted with                           \\ \hline
$N(i)$       &  \begin{tabular}[c]{@{}c@{}} The set of social friends who user $u_i$ \\ directly connected with  \end{tabular}                           \\ \hline
$B(j)$       &  The set of users who have interacted the item $v_j$             \\ \hline
$\mathbf{h}_{i}^{I}$         & \begin{tabular}[c]{@{}c@{}}The item-space user latent factor from\\ item set ${C(i)}$ of user $u_i$   \end{tabular}                            \\ \hline
$\mathbf{h}_{i}^{S}$         & \begin{tabular}[c]{@{}c@{}}The social-space user latent factor from\\ the social friends $N(i)$  of user $u_i$   \end{tabular}                             \\ \hline
$\mathbf{h}_{i}$        & \begin{tabular}[c]{@{}c@{}}The user latent factor of  user $u_i$, combining \\ from item space $\mathbf{h}_{i}^{I}$ and social space $\mathbf{h}_{i}^{S}$ \end{tabular}                          \\ \hline
$\mathbf{x}_{ia}$    &  \begin{tabular}[c]{@{}c@{}}The opinion-aware interaction representation \\ of item $v_a$ for user $u_i$   \end{tabular}    \\ \hline
$\mathbf{f}_{jt}$    &  \begin{tabular}[c]{@{}c@{}}The opinion-aware interaction representation \\ of user $u_t$ for item $v_j$   \end{tabular}  \\ \hline
$\mathbf{z}_{j}$                & The item latent factor of item $v_j$                             \\ \hline
$\alpha _{ia}$           &   \begin{tabular}[c]{@{}c@{}}The item attention of item $v_a$ in \\contributing to $\mathbf{h}_{i}^I$   \end{tabular}                              \\ \hline
$\beta _{io}$           &   \begin{tabular}[c]{@{}c@{}}The social attention of neighboring user $u_o$ in \\contributing to $\mathbf{h}_{i}^S$   \end{tabular}                              \\ \hline
$\mu _{jt}$           &   \begin{tabular}[c]{@{}c@{}}The user attention of user $u_t$ in \\contributing to $\mathbf{z}_{j}$      \end{tabular}                              \\ \hline
$r'_{ij}$        & The predicted rating value of item $v_j$ by user $u_i$           \\ \hline
$\oplus$     &  The concatenation operator of two vectors                            \\ \hline
$\mathrm{T}$ & The user-user social graph \\ \hline
$\mathbf{R} $  & The user-item rating matrix (user-item graph)\\ \hline
$\mathbf{W}, \mathbf{b}$ & The weight and bias in neural network \\
\end{tabular}
\end{table}

\subsection{Definitions and Notations}

Let $U= \left \{ u_1, u_2, ..., u_n \right \}$ and $ V = \left \{ v_1, v_2, ..., v_m \right \} $ be the sets of users and items respectively, where $n$ is the number of users, and $m$ is the number of items. We assume that $\mathbf{R} \in \mathbb{R} ^{n \times m} $ is the user-item rating matrix, which is also called the user-item graph. If $u_i$ gives a rating to $v_j$, $r_{ij}$ is the rating score, otherwise we employ $0$ to represent the unknown rating from $u_i$ to $v_j$, i.e., $r_{ij} = 0$. The observed rating score $r_{ij}$  can be seen as user $u_i$'s opinion on the item $v_j$. Let $\mathcal{O} = \left \{ \left \langle u_i,v_j \right \rangle  | r_{ij} \neq 0 \right \}$ be the set of known ratings and $\mathcal{T} = \left \{ \left \langle u_i,v_j \right \rangle  | r_{ij}  =  0 \right \}$ be the set of unknown ratings. Let $N(i)$ be the set of users whom $u_i$ directly connected with, $C(i)$ be the set of items which $u_i$ have interacted with, and $B(j)$ be the set of users who have interacted with $v_j$.  In addition, users can establish social relations to each other. We use $\mathrm{T} \in \mathbb{R} ^{n \times n}$ to denote the user-user social graph, where $\mathrm{T}_{ij} = 1$ if $u_j$ has a relation to $u_i$ and zero otherwise. Given the user-item graph $\mathbf{R}$ and social graph $\mathrm{T}$, we aim to predict the missing rating value in $\mathbf{R}$. Following~\cite{He2017NCF}, we use an embedding vector $\mathbf{p}_i \in \mathbb{R} ^{d}$ to denote a user $u_i$ and an embedding vector $\mathbf{q}_j \in \mathbb{R} ^{d}$ to represent an item $v_j$, where $d$ is the length of embedding vector. More details will be provided about these embedding vectors in the following subsections.  The mathematical notations used in this paper are summarized in Table~\ref{tab:notation}.


\subsection{An Overview of the Proposed Framework}

The architecture of the proposed model is shown in Figure~\ref{fig:GraphRec}. The model consists of three components: user modeling, item modeling, and rating prediction. The first component is user modeling, which is to learn latent factors of users. As data in social recommender systems includes two different graphs, i.e., a social graph and a user-item graph, we are provided with a great opportunity to learn user representations from different perspectives. Therefore, two aggregations are introduced to respectively process these two different graphs. One is item aggregation, which can be utilized to understand users via interactions between users and items in the user-item graph (or item-space). The other is social aggregation, the relationship between users in the social graph, which can help model users from the social perspective (or social-space). Then, it is intuitive to obtain user latent factors by combining information from both item space and social space. The second component is item modeling, which is to learn latent factors of items. In order to consider both interactions and opinions in the user-item graph, we introduce user aggregation, which is to aggregate users' opinions in item modeling. The third component is to learn model parameters via prediction by integrating user and item modeling components. Next, we will detail each model component.

\subsection{User Modeling}
\label{sec:User_Modeling}

User modeling aims to learn user latent factors, denoted as $\mathbf{h}_{i} \in \mathbb{R} ^{d}$ for user $u_i$. The challenge is how to inherently combine the user-item graph and social graph. To address this challenge, we first use two types of aggregation to learn factors from two graphs, as shown in the left part in Figure~\ref{fig:GraphRec}.  The first aggregation, denoted as item aggregation, is utilized to learn item-space user latent factor $\mathbf{h}_{i}^{I} \in \mathbb{R} ^{d} $ from the user-item graph. The second aggregation is social aggregation where social-space user latent factor $\mathbf{h}^{S}_{i} \in \mathbb{R} ^{d}$ is learned from the social graph.  Then, these two factors are combined together to form the final user latent factors $\mathbf{h}_{i}$. Next, we will introduce item aggregation, social aggregation and how to combine user latent factors from both item-space and social-space. 

{\bf Item Aggregation.} As user-item graph contains not only interactions between users and items but also users' opinions (or rating scores) on items, we provide a principled approach to jointly capture interactions and opinions in the user-item graph for learning item-space user latent factors $\mathbf{h}_{i}^{I}$, which is used to model user latent factor via interactions in the user-item graph.
 
The purpose of item aggregation is to learn item-space user latent factor $\mathbf{h}_{i}^{I}$ by considering items a user $u_i$ has interacted with and users' opinions on these items. To mathematically represent this aggregation, we use the following function as:
\begin{align}
\mathbf{h}^{I}_{i} = \sigma ( \mathbf{W}  \cdot  Aggre_{items} (\left \{ \mathbf{x}_{ia}, \forall a \in C(i) \right \}) + \mathbf{b})
\end{align}
where $C(i)$ is the set of items user $u_i$ has interacted with (or $u_i$'s neighbors in the user-item graph), $\mathbf{x}_{ia}$ is a representation vector to denote opinion-aware interaction between $u_i$ and an item $v_a$, and $Aggre_{items}$ is the items aggregation function. In addition, $\sigma$ denotes non-linear activation function (i.e., a rectified linear unit), and $\mathbf{W}$ and $\mathbf{b}$ are the weight and bias of a neural network. Next we will discuss how to define opinion-aware interaction representation $\mathbf{x}_{ia}$ and the aggregation function $Aggre_{items}$. 

A user can express his/her opinions (or rating scores), denoted as $r$, to items during user-item interactions. These opinions on items can capture users' preferences on items, which can help model item-space user latent factors. To model opinions, for each type of opinions $r$, we introduce an opinion embedding vector $\mathbf{e}_r\in \mathbb{R}^{d}$ that denotes each opinion $r$ as a dense vector representation. For example, in a 5-star rating system, for each $r \in \left \{ 1,2,3,4,5 \right \}$, we introduce an embedding vector $\mathbf{e}_r$. For an interaction between user $u_i$ and item $v_a$ with opinion $r$, we model opinion-aware interaction representation $\mathbf{x}_{ia}$ as a combination of item embedding $\mathbf{q}_a$ and opinion embedding $\mathbf{e}_r$ via a Multi-Layer Perceptron (MLP). It can be denoted as $g_v$ to fuse the interaction information with the opinion information as shown in Figure~\ref{fig:GraphRec}. The MLP takes the concatenation of item embedding $\mathbf{q}_{a}$ and its opinion embedding $\mathbf{e}_{r}$ as input. The output of MLP is the opinion-aware representation of the interaction between $u_i$ and $v_a$, $\mathbf{x}_{ia}$, as follows:
\begin{align}
\mathbf{x}_{ia} = g_v(\left [  \mathbf{q}_{a} \oplus  \mathbf{e}_{r} \right ])
\end{align}
where $\oplus$ denotes the concatenation operation between two vectors.


One popular aggregation function for $Aggre_{items}$ is the mean operator where we take the element-wise mean of the vectors in $\left \{ \mathbf{x}_{ia}, \forall a \in C(i) \right \}$. This mean-based aggregator is a linear approximation of a localized spectral convolution~\cite{kipf2017semi}, as the following function:
\begin{align}
\mathbf{h}^{I}_{i}  = \sigma (  \mathbf{W} \cdot   \left \{  \sum_{ a \in C(i)} \alpha _i \mathbf{x}_{ia} \right \}  + \mathbf{b})
\end{align}
where $\alpha _i$ is fixed to $\frac{1}{\left |C(i)  \right |} $ for all items in the mean-based aggregator. It assumes that all interactions contribute equally to understand the user $u_i$. However, this may not be optimal, due to the fact that the influence of interactions on users may vary dramatically. Hence, we should allow interactions to contribute differently to a user's latent factor by assigning each interaction a weight.

To alleviate the limitation of mean-based aggregator, inspired by attention mechanisms~\cite{Chen2018Neural, yang2016hierarchical}, an intuitive solution is to tweak $\alpha_{i}$  to be aware of the target user $u_i$, i.e., assigning an individualized weight for each $(v_a, u_i)$ pair,
\begin{align}
\mathbf{h}^{I}_{i} =  \sigma ( \mathbf{W} \cdot \left \{  \sum_{ a \in C(i)} \alpha _{ia} \mathbf{x}_{ia} \right \}  + \mathbf{b})\label{ItemAggregation}
\end{align}
where $\alpha _{ia}$ denotes the attention weight of the interaction with $v_a$ in contributing to user $u_i$'s item-space latent factor when characterizing user $u_i$'s preference from the interaction history $C(i)$. Specially, we parameterize the \emph{item attention} $\alpha _{ia}$ with a two-layer neural network, which we call as the \emph{attention network}.  The input to the attention network is the opinion-aware representation $\mathbf{x}_{ia}$ of the interaction and the target user $u_i$'s embedding $\mathbf{p}_i$. Formally, the attention network is defined as,
\begin{align}
\alpha _{ia}^* &=\mathbf{w}_2^T \cdot \sigma(\mathbf{W}_{1} \cdot [\mathbf{x}_{ia} \oplus  \mathbf{p}_i] + \mathbf{b}_1) + b_2
\end{align}

The final attention weights are obtained by normalizing the above attentive scores using Softmax function, which can be interpreted as the contribution of the interaction to the item-space user latent factor of user $u_i$ as:
\begin{align}
\alpha _{ia} &= \frac{exp(\alpha _{ia}^*)}{\sum_{a \in C(i)} exp(\alpha _{ia}^*)}
\end{align}

{\bf Social Aggregation.} Due to the social correlation theories~\cite{mcpherson2001birds,marsden1993network},  a user's preference is similar to or influenced by his/her directly connected social friends. We should incorporate social information to further model user latent factors.
Meanwhile, tie strengths between users can further influence users' behaviors from the social graph. In other words, the learning of social-space user latent factors should consider heterogeneous strengths of social relations. Therefore, we introduce an attention mechanism to select social friends that are representative to characterize users social information and then aggregate their information.

In order to represent user latent factors from this social perspective, we propose social-space user latent factors, which is to aggregate the item-space user latent factors of neighboring users from the social graph. Specially, the social-space user latent factor of $u_i$, $\mathbf{h}_{i}^S$, is to aggregate the item-space user latent factors of users in $u_i$'s neighbors ${N(i)}$, as the follows:
\begin{align}
\mathbf{h}_{i}^S =  \sigma( \mathbf{W}  \cdot Aggre_{neigbhors}(\left \{ \mathbf{h}_o^I, \forall o \in N(i) \right \}) + \mathbf{b} )
\end{align}
where $Aggre_{neigbhors}$ denotes the aggregation function on user's neighbors.

One natural aggregation function for $Aggre_{neigbhors}$ is also the mean operator which take the element-wise mean of the vectors in $\left \{ \mathbf{h}_o^I, \forall o \in N(i) \right \}$, as the following function:
\begin{align}
\mathbf{h}_{i}^S =  \sigma( \mathbf{W}  \cdot  \left \{ \sum_{ o \in N(i)} \beta  _i \mathbf{h}_o^I \right \} + \mathbf{b})
\end{align}
where $\beta _i$ is fixed to $\frac{1}{\left |N(i)  \right |} $ for all neighbors for the mean-based aggregator. It assumes that all neighbors contribute equally to the representation of user $u_i$. However, as mentioned before, strong and weak ties are mixed together in a social network, and users are likely to share more similar tastes with strong ties than weak ties. Thus, we perform an attention mechanism with a two-layer neural network to extract these users that are important to influence $u_i$, and model their tie strengths, by relating \emph{social attention} $\beta_{io}$ with $\mathbf{h}_o^I$ and the target user embedding $\mathbf{p}_i$, as below,
\begin{align}
\mathbf{h}_{i}^S  &=  \sigma (  \mathbf{W}  \cdot  \left \{ \sum_{ o \in N(i)} \beta _{io} \mathbf{h}_o^I  \right \} + \mathbf{b} ) \\
\beta _{io}^* &=\mathbf{w}_2^T \cdot \sigma(\mathbf{W}_{1} \cdot[\mathbf{h}_o^I  \oplus  \mathbf{p}_i] + \mathbf{b}_1) + b_2 \\
\beta _{io} &= \frac{exp(\beta_{io}^*)}{\sum_{o \in N(i)} exp(\beta_{io}^*)}
\end{align}
\noindent where the $\beta _{io}$  can be seen as the strengths between users.

{\bf Learning User Latent Factor.} In order to learn better user latent factors, item-space user latent factors and social-space user latent factors are needed to be considered together, since the social graph and the user-item graph provide information about users from different perspectives. We propose to combine these two latent factors to the final user latent factor via a standard MLP where the item-space user latent factor $\mathbf{h}_i^I $ and the social-space user latent factor $\mathbf{h}_i^S $ are concatenated before feeding into MLP. Formally, the user latent factor $\mathbf{h}_{i}$ is defined as,
\begin{align}
\mathbf{c}_{1} &= \left [ \mathbf{h}_i^I \oplus   \mathbf{h}_i^S \right ]\\
\mathbf{c}_{2} &= \sigma (\mathbf{W}_{2} \cdot \mathbf{c}_{1} + \mathbf{b}_{2})    \\
&...\nonumber\\
\mathbf{h}_{i} &= \sigma (\mathbf{W}_{l} \cdot \mathbf{c}_{l-1} + \mathbf{b}_{l})
\end{align}
where $l$ is the index of a hidden layer. 

\subsection{Item Modeling}

As shown in the right part of Figure~\ref{fig:GraphRec}, item modeling is used to learn item latent factor, denoted as $\mathbf{z}_{j}$, for the item $v_j$ by user aggregation. Items are associated with the user-item graph, which contains interactions as well as user's opinions. Therefore, interactions and opinions in the user-item graph should be jointly captured to further learn item latent factors.

{\bf User Aggregation.} Likewise, we use a similar method as learning item-space user latent factors via item aggregation. For each item $v_j$, we need to aggregate information from the set of users who have interacted with $v_j$, denoted as $B(j)$. 

Even for the same item, users might express different opinions during user-item interactions. These opinions from different users can capture the characteristics of the same item in different ways provided by users, which can help model item latent factors. For an interaction from $u_t$ to $v_j$ with opinion $r$, we introduce an opinion-aware interaction user representation $\mathbf{f}_{jt}$, which is obtained from the basic user embedding $\mathbf{p}_t$ and opinion embedding $\mathbf{e}_r$ via a MLP, denoted as $g_u$. $g_u$ is to fuse the interaction information with the opinion information, as shown in Figure~\ref{fig:GraphRec}:
\begin{align}
\mathbf{f}_{jt} = g_u(\left [  \mathbf{p}_{t} \oplus  \mathbf{e}_{r} \right ])
\end{align}

Then, to learn item latent factor $\mathbf{z}_j$, we also propose to aggregate opinion-aware interaction representation of users in $B(j)$ for item $v_j$. The users aggregation function is denoted as $Aggre_{users}$, which is to aggregate opinion-aware interaction representation of users in $\left \{ \mathbf{f}_{jt}, \forall t \in B(j) \right \}$ as:
\begin{align}
\mathbf{z}_{j} =  \sigma( \mathbf{W}  \cdot   Aggre_{users}(\left \{ \mathbf{f}_{jt}, \forall t \in B(j) \right \}) + \mathbf{b})
\end{align}

In addition, we introduce an attention mechanism to differentiate the importance weight $\mu _{jt}$  of users with a two-layer neural attention network, taking $\mathbf{f}_{jt}$ and $\mathbf{q}_j$ as the input,
\begin{align}
\mathbf{z}_{j} &=  \sigma ( \mathbf{W}  \cdot  \left \{ \sum_{ t \in B(j)} \mu _{jt} \mathbf{f}_{jt} \right \} + \mathbf{b}) \\
\mu_{jt}^* &=\mathbf{w}_2^T \cdot \sigma(\mathbf{W}_{1} \cdot [\mathbf{f}_{jt} \oplus  \mathbf{q}_j] + \mathbf{b}_1) + b_2 \\
\mu _{jt} &= \frac{exp(\mu_{jt}^*)}{\sum_{t \in B_{(j)}} exp(\mu_{jt}^*)}
\end{align}
This \emph{user attention} $\mu _{jt}$ is to capture heterogeneous influence from user-item interactions on learning item latent factor. 

\subsection{Rating Prediction}

In this subsection, we will design recommendation tasks to learn model parameters. There are various recommendation tasks such as item ranking and rating prediction. In this work, we apply the proposed GraphRec model for the recommendation task of rating prediction. With the latent factors of users and items (i.e., $\mathbf{h}_{i}$ and $\mathbf{z}_{j}$), we can first concatenate them $\left [ \mathbf{h}_{i} \oplus  \mathbf{z}_{j} \right ]$ and then feed it into MLP for rating prediction as:
\begin{align}
\mathbf{g}_{1} &= \left [ \mathbf{h}_{i} \oplus  \mathbf{z}_{j} \right ]\\
\mathbf{g}_{2} &= \sigma (\mathbf{W}_{2} \cdot  \mathbf{g}_{1} + \mathbf{b}_{2})    \\
&...\nonumber\\
\mathbf{g}_{l-1} &= \sigma (\mathbf{W}_{l} \cdot \mathbf{g}_{l-1} + \mathbf{b}_{l}) \\
{r}'_{ij} &=\mathbf{w}^{T} \cdot \mathbf{g}_{l-1}
\end{align}
where $l$ is the index of a hidden layer, and ${r}'_{ij}$ is the predicted rating from $u_i$ to $v_j$.

\subsection{Model Training}

To estimate model parameters of GraphRec, we need to specify an objective function to optimize. Since the task we focus on in this work is rating prediction, a commonly used objective function is formulated as,

\begin{align}
\label{equ:loss}
Loss = \frac{1}{ 2 \left | \mathcal{O} \right | }  \sum _{i,j \in \mathcal{O}} (r'_{ij} - r_{ij})^2
\end{align}
\noindent where $|\mathcal{O}|$ is the number of observed ratings , and $r_{ij}$ is the ground truth rating assigned by the user $i$ on the item $j$.

To optimize the objective function, we adopt the RMSprop~\cite{tieleman2012lecture} as the optimizer in our implementation, rather than the vanilla SGD. At each time, it randomly selects a training instance and updates each model parameter towards the direction of its negative gradient. There are three embedding in our model, including item embedding $\mathbf{q}_j$, user embedding $\mathbf{p}_i$, and opinion embedding $\mathbf{e}_r$. They are randomly initialized and jointly learned during the training stage.  We do not use one-hot vectors to represent each user and item, since the raw features are very large and highly sparse. By embedding high-dimensional sparse features into a low-dimensional latent space, the model can be easy to train~\cite{He2017NCF}. Opinion embedding matrix $\mathbf{e}$ depends on the rating scale of the system. For example, for a 5-star rating system,  opinion embedding matrix $\mathbf{e}$ contains 5 different embedding vectors to denote scores in $\left \{ 1,2,3,4,5 \right \}$. Overfitting is a perpetual problem in optimizing deep neural network models. To alleviate this issue, the dropout strategy~\citep{srivastava2014dropout} has been applied to our model. The idea of dropout is to randomly drop some neurons during the training process. When updating parameters, only part of them will be updated. Moreover, as dropout is disabled during testing, the whole network is used for prediction.

\section{Experiment}
\label{sec:Experiments}

\subsection{Experimental Settings}
\subsubsection{Datasets}
In our experiments, we choose two representative datasets Ciao and Epinions\footnote{http://www.cse.msu.edu/$\sim$tangjili/index.html}, which are taken from popular social networking websites Ciao (http://www.ciao.co.uk) and Epinions (www.epinions.com). Each social networking service allows users to rate items, browse/write reviews, and add friends to their `Circle of Trust'. Hence, they provide a large amount of rating information and social information. The ratings scale is from 1 to 5. We randomly initialize opinion embedding with 5 different embedding vectors based on 5 scores in $\left \{ 1,2,3,4,5 \right \}$.  The statistics of these two datasets are presented in~\tablename ~\ref{tab:dataset}.


\begin{table}[htbp]
\centering
\caption{Statistics of the datasets}
\label{tab:dataset}
\begin{tabular}{l|c|c}
\hline
Dataset               & Ciao  &Epinions \\ \hline
\# of Users           & 7,317   &18,088 \\ \hline
\# of Items           & 10,4975  &261,649 \\ \hline
\# of Ratings          & 283,319  &764,352\\ \hline
\# of Density (Ratings)          & 0.0368\% &0.0161\%  \\ \hline \hline
\# of Social Connections & 111,781 &355,813 \\ \hline
\# of Density (Social Relations)  & 0.2087\% &0.1087\% \\ \hline
\end{tabular}
\vskip -0.1in
\end{table}

\subsubsection{Evaluation Metrics}
In order to evaluate the quality of the recommendation algorithms, two popular metrics are adopted to evaluate the predictive accuracy, namely Mean Absolute Error (MAE) and Root Mean Square Error (RMSE)~\cite{wang2018exploring}. Smaller values of MAE and RMSE indicate better predictive accuracy. Note that small improvement in RMSE or MAE terms can have a significant impact on the quality of the top-few recommendations~\cite{koren2008factorization}.

\subsubsection{Baselines}

To evaluate the performance, we compared our GraphRec with three groups of methods including traditional recommender systems, traditional social recommender systems, and deep neural network based recommender systems. For each group, we select representative baselines and below we will detail them.

\begin{itemize}
  \item \textbf{PMF}~\cite{salakhutdinov2007probabilistic}: \textbf{P}robabilistic \textbf{M}atrix \textbf{F}actorization utilizes user-item rating matrix only and models latent factors of users and items by Gaussian distributions.
  \item \textbf{SoRec}~\cite{ma2008sorec}: \textbf{So}cial \textbf{Rec}ommendation performs co-factorization on the user-item rating matrix and user-user social relations matrix.
  \item \textbf{SoReg}~\cite{ma2011recommender}: \textbf{So}cial \textbf{Reg}ularization models social network information as regularization terms to constrain the matrix factorization framework.
  \item \textbf{SocialMF}~\cite{jamali2010matrix}: It considers the trust information and propagation of trust information into the matrix factorization model for recommender systems.
  \item \textbf{TrustMF}~\cite{yang2017social}: This method adopts matrix factorization technique that maps users into two low-dimensional spaces: truster space and trustee space, by factorizing trust networks according to the directional property of trust.
  \item \textbf{NeuMF}~\cite{He2017NCF}: This method is a state-of-the-art matrix factorization model with neural network architecture. The original implementation is for recommendation ranking task and we adjust its loss to the squared loss for rating prediction.
  \item \textbf{DeepSoR}~\cite{DeepSoR2018}: This model employs a deep neural network to learn representations of each user from social relations, and to integrate into probabilistic matrix factorization for rating prediction.
  \item \textbf{GCMC+SN}~\cite{berg2017graph}: This model is a state-of-the-art recommender system with graph neural network architecture. In order to incorporate social network information into GCMC, we utilize the $node2vec$~\cite{grover2016node2vec} to generate user embedding as user side information, instead of using the raw feature social connections ($\mathrm{T} \in \mathbb{R} ^{n \times n}$) directly. The reason is that the raw feature input vectors is highly sparse and high-dimensional. Using the network embedding techniques can help compress the raw input feature vector to a low-dimensional and dense vector, then the model can be easy to train.
\end{itemize}

PMF and NeuMF are pure collaborative filtering model without social network information for rating prediction, while the others are social recommendation. Besides, we compared GraphRec with two state-of-the-art neural network based social recommender systems, i.e., DeepSoR and GCMC+SN. 

\begin{table*}[htbp]
\centering
\caption{Performance comparison of different recommender systems}
\label{tab:baselines_results}
\begin{tabular}{|c|c|c|c|c|c|c|c|c|c|c|}
\hline
\multirow{2}{*}{Training}                                                   & \multirow{2}{*}{Metrics} & \multicolumn{9}{c|}{Algorithms}                                             \\ \cline{3-11}
                                                                           &                          & PMF    & SoRec  & SoReg  & SocialMF & TrustMF & NeuMF  & DeepSoR  & GCMC+SN  & \textbf{GraphRec} \\ \hline
\multirow{2}{*}{\begin{tabular}[c]{@{}c@{}}Ciao\\ (60\%)\end{tabular}}     & MAE                      & 0.952  & 0.8489 & 0.8987 & 0.8353   & 0.7681   & 0.8251 & 0.7813  & 0.7697  & 0.7540  \\ \cline{2-11}
                                                                           & RMSE                     & 1.1967 & 1.0738 & 1.0947 & 1.0592   & 1.0543   & 1.0824 & 1.0437  & 1.0221  & 1.0093  \\ \hline
\multirow{2}{*}{\begin{tabular}[c]{@{}c@{}}Ciao\\ (80\%)\end{tabular}}     & MAE                      & 0.9021 & 0.8410 & 0.8611 & 0.8270   & 0.7690   & 0.8062 & 0.7739  & 0.7526  & 0.7387  \\ \cline{2-11}
                                                                           & RMSE                     & 1.1238 & 1.0652 & 1.0848 & 1.0501   & 1.0479   & 1.0617 & 1.0316  & 0.9931  & 0.9794  \\ \hline \hline
\multirow{2}{*}{\begin{tabular}[c]{@{}c@{}}Epinions\\ (60\%)\end{tabular}} & MAE                      & 1.0211 & 0.9086 & 0.9412 & 0.8965   & 0.8550   & 0.9097 & 0.8520   & 0.8602 & 0.8441   \\ \cline{2-11}
                                                                           & RMSE                     & 1.2739 & 1.1563 & 1.1936 & 1.1410   & 1.1505   & 1.1645 & 1.1135  & 1.1004  & 1.0878    \\ \hline
\multirow{2}{*}{\begin{tabular}[c]{@{}c@{}}Epinions\\ (80\%)\end{tabular}} & MAE                      & 0.9952 & 0.8961 & 0.9119 & 0.8837   & 0.8410   & 0.9072 & 0.8383  & 0.8590  & 0.8168  \\ \cline{2-11}
                                                                           & RMSE                     & 1.2128 & 1.1437 & 1.1703 & 1.1328   & 1.1395   & 1.1476 & 1.0972  & 1.0711  & 1.0631  \\ \hline
\end{tabular}
\end{table*}


\subsubsection{Parameter Settings}
We implemented our proposed method on the basis of Pytorch\footnote{https://pytorch.org/}, a well-known Python library for neural networks.
For each dataset, we used $x$\% as a training set to learning parameters, $(1-x\%)/2$ as a validation set to tune hyper-parameters, and $(1-x\%)/2$ as a testing set for the final performance comparison, where $x$ was varied as $\left \{ 80\%, 60\% \right \}$. For the embedding size $d$, we tested the  value of [ 8, 16, 32, 64, 128, 256 ].  The batch size and learning rate was searched in [ 32, 64, 128, 512 ] and [ 0.0005, 0.001, 0.005, 0.01, 0.05, 0.1 ], respectively. Moreover, we empirically set the size of the hidden layer the same as the embedding size and the activation function as ReLU.  Without special mention, we employed three hidden layers for all the neural components. The early stopping strategy was performed, where we stopped training if the RMSE on validation set increased for 5 successive
epochs. For all neural network methods, we randomly initialized model parameters with a Gaussian distribution, where the mean and standard deviation is 0 and 0.1, respectively. The parameters for the baseline algorithms were initialized as in the corresponding papers and were then carefully tuned to achieve optimal performance.

\subsection{Performance Comparison of Recommender Systems}
We first compare the recommendation performance of all methods. Table~\ref{tab:baselines_results} shows the overall rating prediction error $w.r.t.$ RMSE and MAE among the recommendation methods on Ciao and Epinions datasets. We have the following main findings:

\begin{itemize}
  \item SoRec, SoReg, SocialMF, and TrustMF always outperform PMF.  All of these methods are based on matrix factorization. SoRec, SoReg, SocialMF, and TrustMF leverage both the rating and social network information; while PMF only uses the rating information. These results support that social network information is complementary to rating information for recommendations.
   \item NeuMF obtains much better performance than PMF.  Both methods only utilize the rating information. However, NeuMF is based on neural network architecture, which suggests the power of neural network models in recommender systems.
   \item DeepSoR and GCMC+SN perform better than SoRec, SoReg, SocialMF, and TrustMF. All of them take advantage of both rating and social network information. However, DeepSoR and GCMC+SN are based on neural network architectures, which further indicate the power of neural network models in recommendations.
   \item Among baselines, GCMC+SN shows quite strong performance. It implies that the GNNs are powerful in representation learning for graph data, since it naturally integrates the node information as well as topological structure.
  \item Our method GraphRec consistently outperforms all the baseline methods. Compared to DeepSoR and GCMC+SN, our model provides advanced model components to integrate rating and social network information. In addition, our model provides a way to consider both interactions and opinions in the user-item graph. We will provide further investigations to better understand the contributions of model components to the proposed framework in the following subsection.
\end{itemize}

To sum up, the comparison results suggest (1) social network information is helpful for recommendations; (2) neural network models can boost recommendation performance and (3) the proposed framework outperforms representative baselines.

\begin{figure*}[t]
\centering
{\subfigure[Ciao-RMSE]
{\includegraphics[width=0.245\linewidth]{{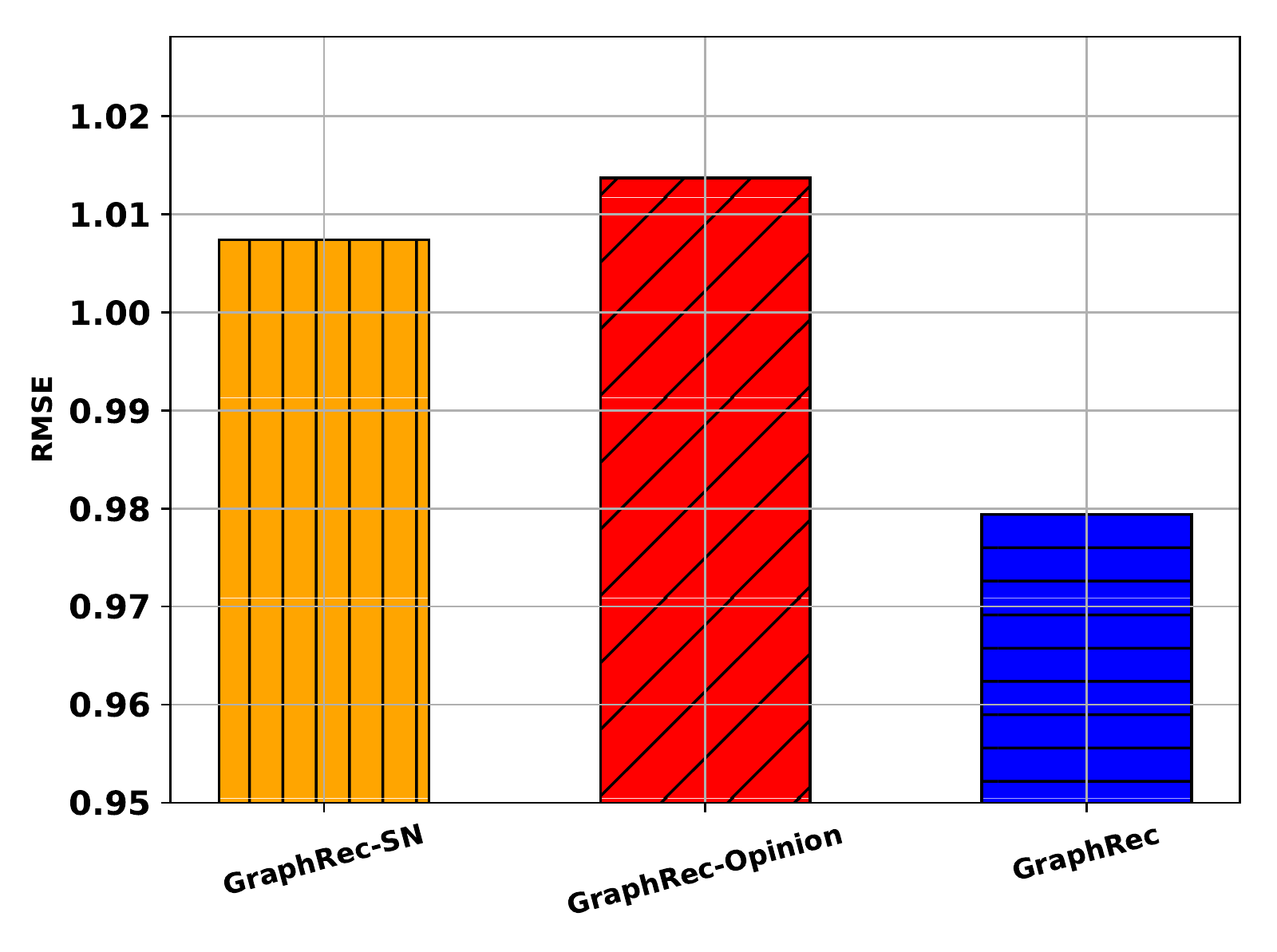}}\label{fig:Ciao_MAE}}}
{\subfigure[Ciao-MAE]
{\includegraphics[width=0.245\linewidth]{{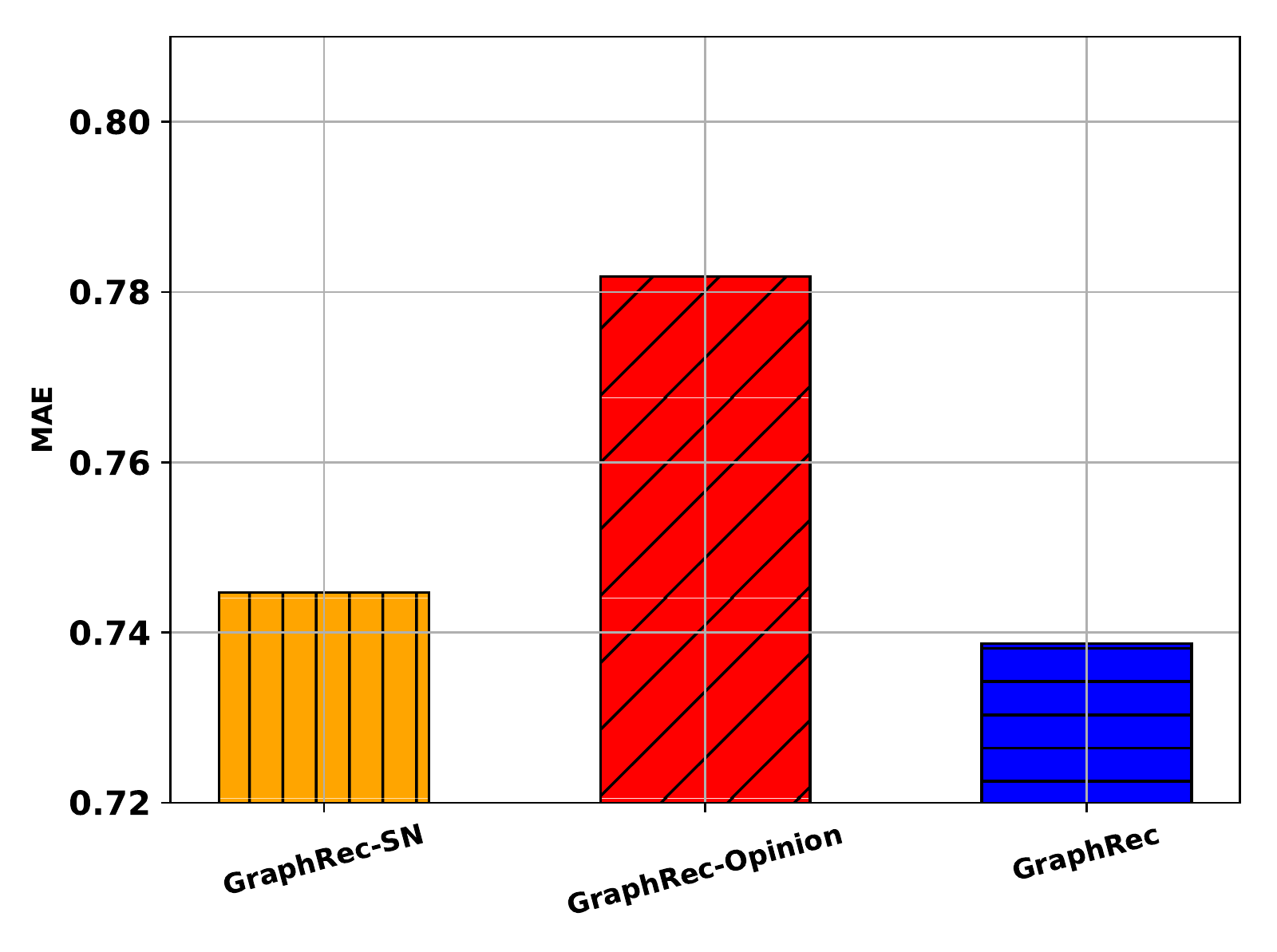}}\label{fig:GraphSR_Ciao_RMSE}}}
{\subfigure[Epinions-RMSE]
{\includegraphics[width=0.245\linewidth]{{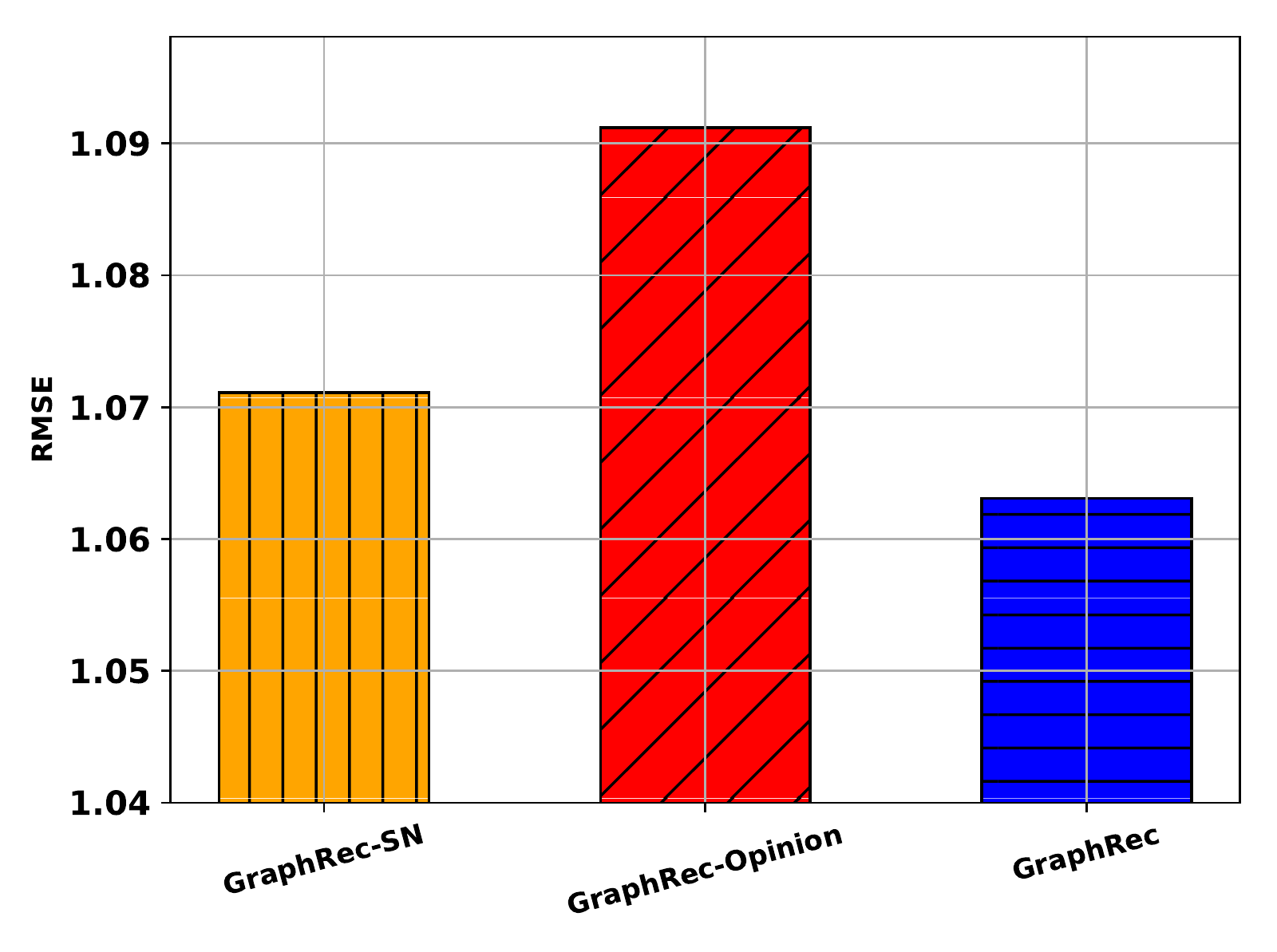}}\label{fig:GraphSR_Epinions_MAE}}}
{\subfigure[Epinions-MAE]
{\includegraphics[width=0.245\linewidth]{{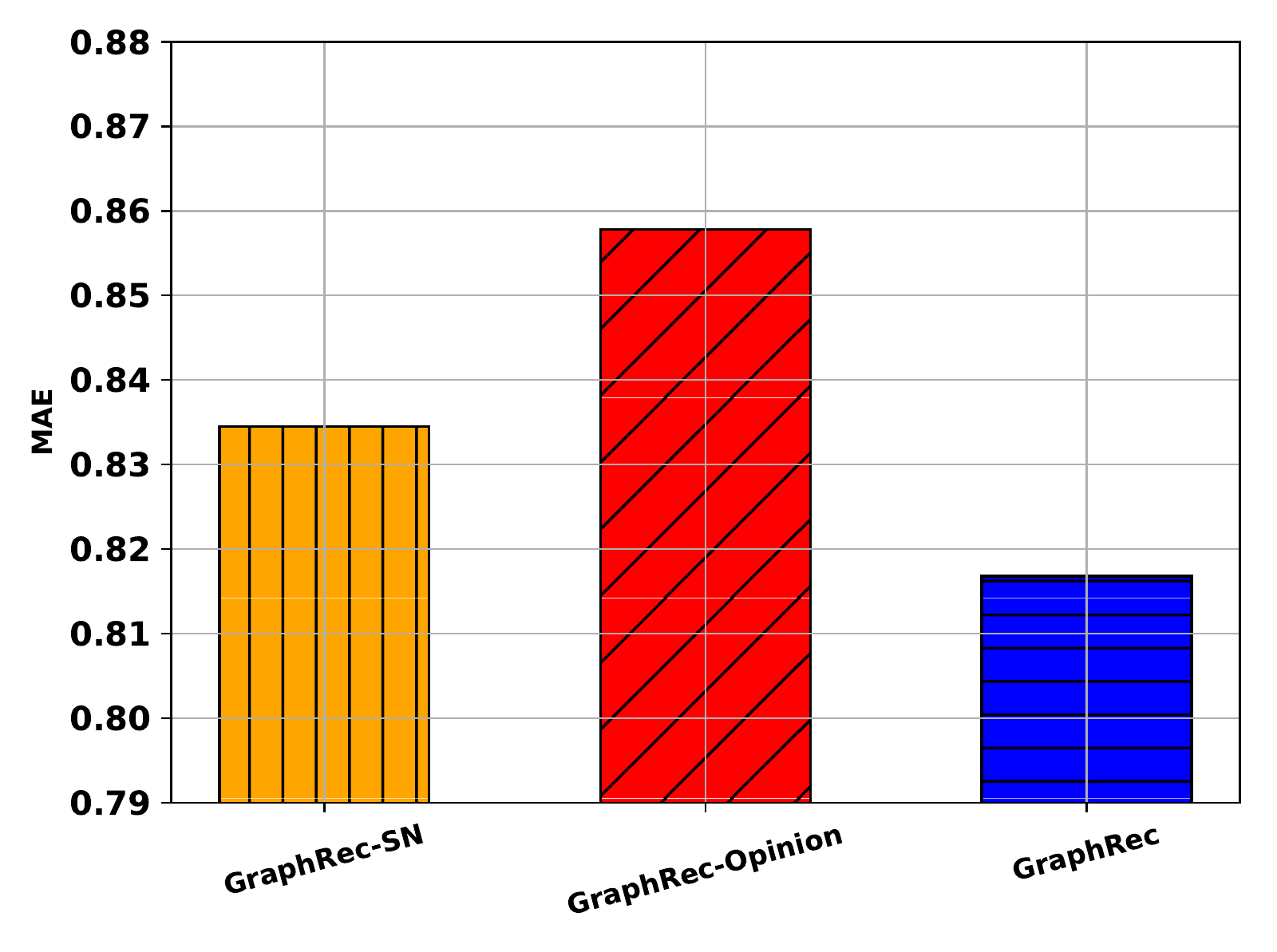}}\label{fig:GraphSR_Epinions_RMSE}}}
\caption{Effect of social network and user opinions on Ciao and Epinions datasets.}\label{fig:GraphSR_sr}
\end{figure*}

\subsection{Model Analysis}

In this subsection, we study the impact of model components and model hyper-parameters. 

\subsubsection{Effect of Social Network and User Opinions}

In the last subsection, we have demonstrated the effectiveness of the proposed framework. The proposed framework provides model components to (1) integrate social network information  and (2) incorporate users' opinions about the interactions with items. To understand the working of GraphRec, we compare GraphRec with its two variants: GraphRec-SN, and GraphRec-Opinion. These two variants are defined in the following:
\begin{itemize}
  \item GraphRec-SN: The social network information of GraphRec is removed . This variant only uses the item-space user latent factor $\mathbf{h}_i^I$ to represent user latent factors $\mathbf{h}_i$; while ignoring the social-space user latent factors $\mathbf{h}_i^S$.
  \item GraphRec-Opinion: For learning item-space user latent factor and item latent factor, the opinion embedding is removed during learning $\mathbf{x}_{ia}$ and $\mathbf{f}_{jt}$. This variant ignores the users' opinion on the user-item interactions.
\end{itemize}

The performance of GraphRec and its variants on Ciao and Epinions are given in Figure~\ref{fig:GraphSR_sr}. From the results, we have the following findings:
\begin{itemize}
  \item {\bf Social Network Information.} We now focus on analyzing the effectiveness of social network information. GraphRec-SN performs worse than GraphRec. It verifies that social network information is important to learn user latent factors and boost the recommendation performance.
  \item {\bf Opinions in Interaction.} We can see that without opinion information, the performance of rating prediction is deteriorated significantly. For example, on average, the relative reduction on Ciao and Epinions is 3.50\% and 2.64\% on RMSE metric, and 5.84\% and 5.02\% on MAE metric, respectively. It justifies our assumption that opinions on user-item interactions have informative information that can help to learn user or item latent factors and improve the performance of recommendation.
\end{itemize}

\begin{figure*}[t]
\centering
{\subfigure[Ciao-RMSE]
{\includegraphics[width=0.245\linewidth]{{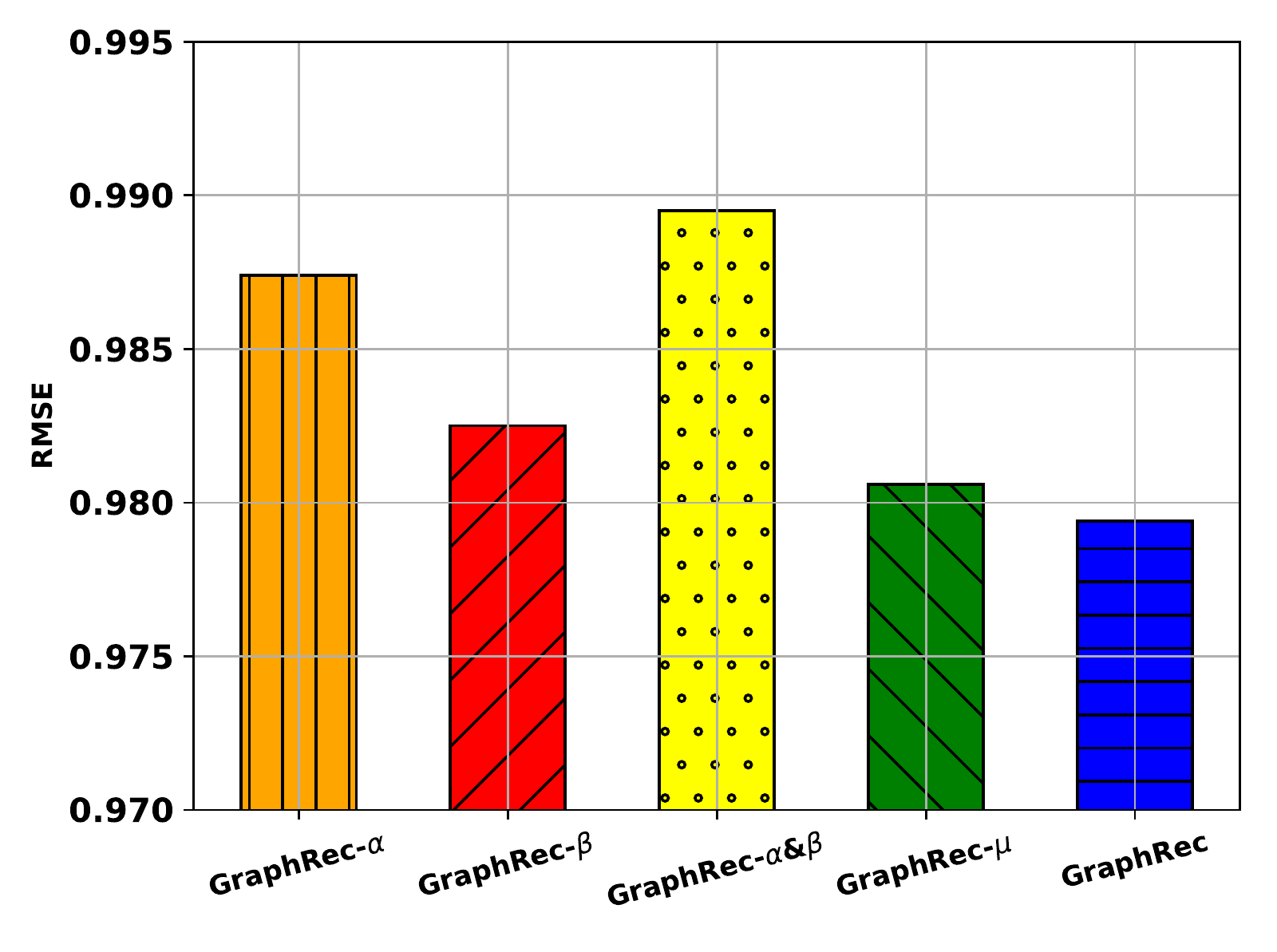}}\label{fig:ciao_att_rmse}}}
{\subfigure[Ciao-MAE]
{\includegraphics[width=0.245\linewidth]{{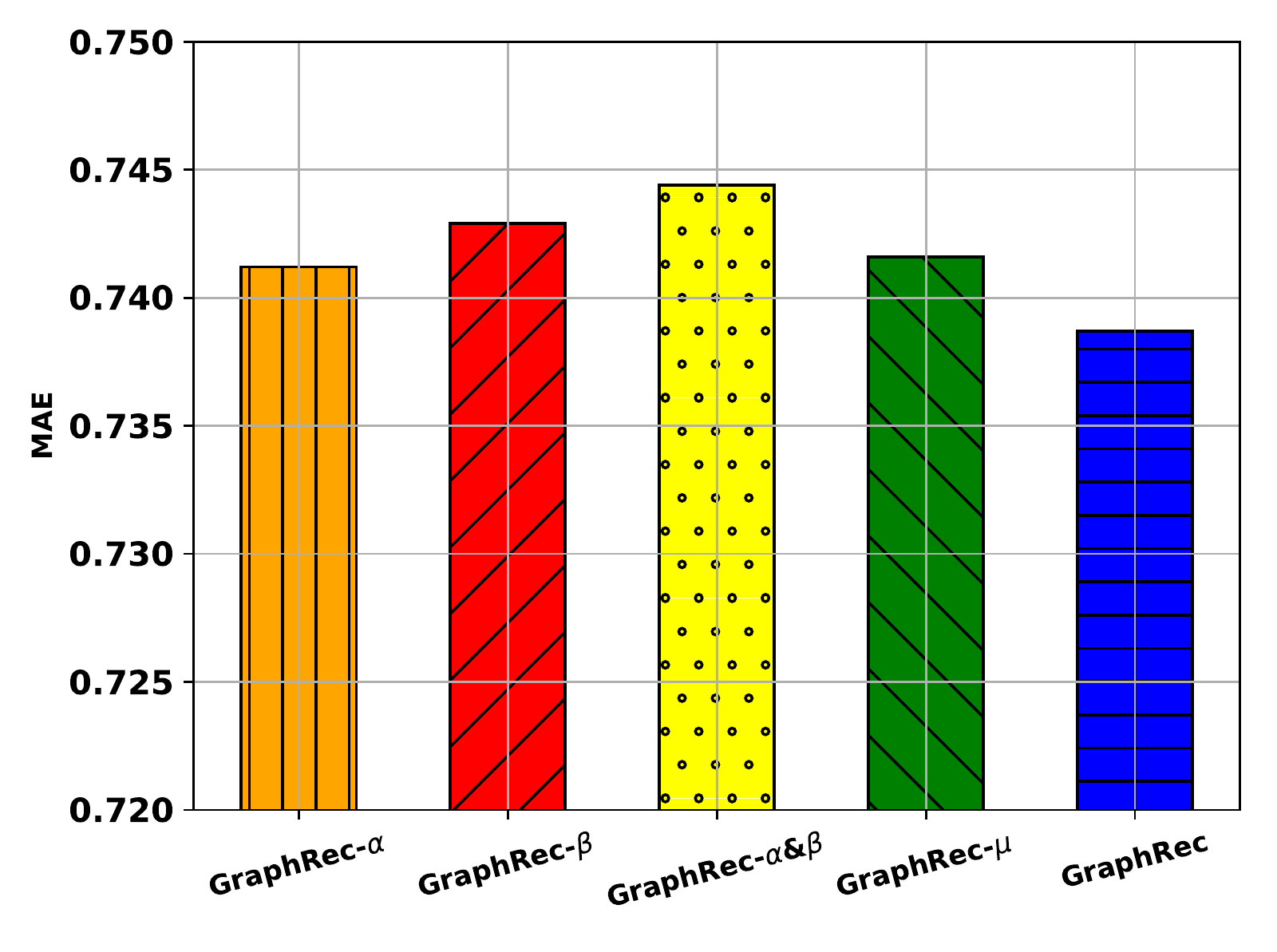}}\label{fig:ciao_att_mae}}}
{\subfigure[Epinions-RMSE]
{\includegraphics[width=0.245\linewidth]{{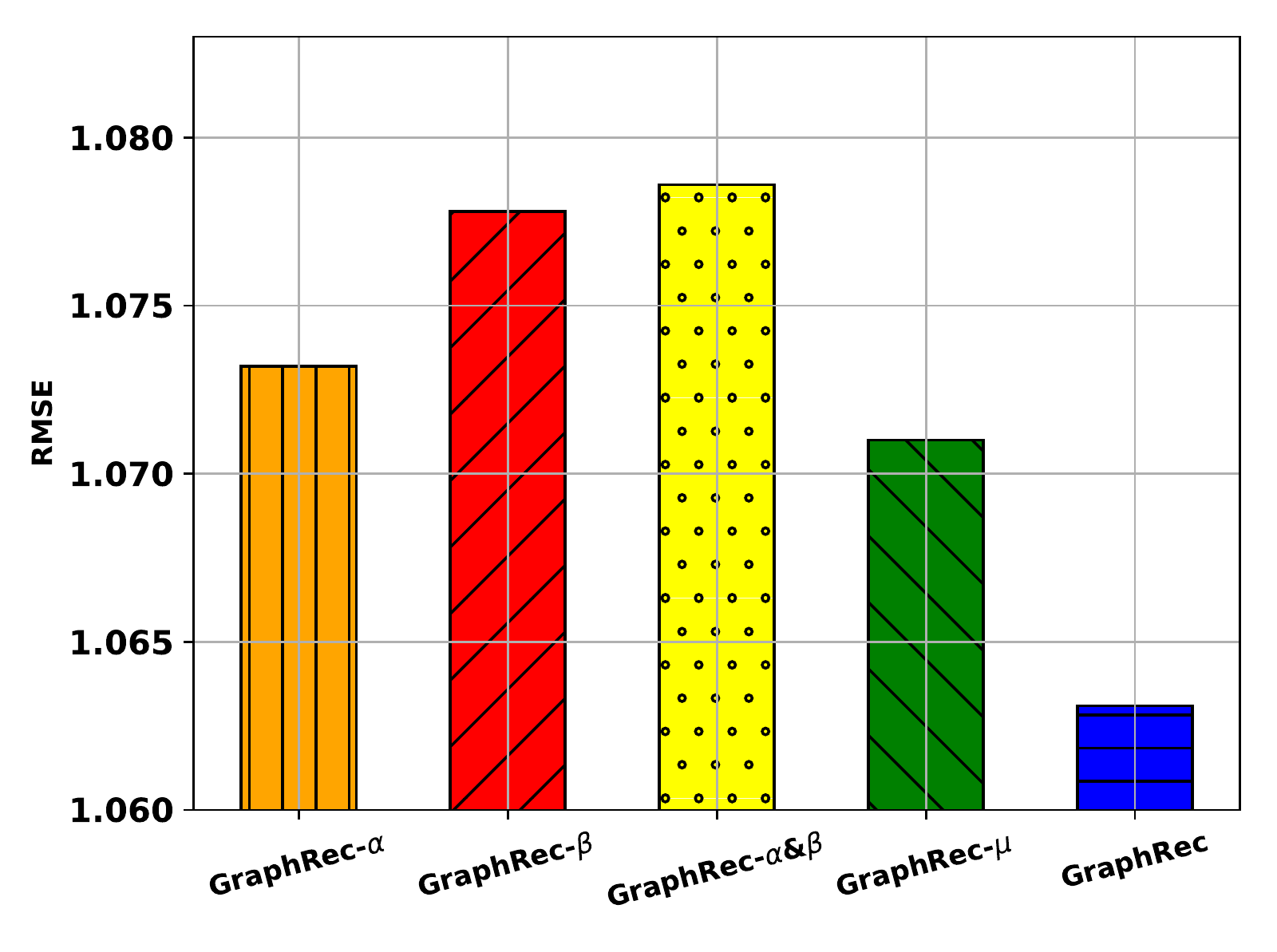}}\label{fig:epinions_att_rmse}}}
{\subfigure[Epinions-MAE]
{\includegraphics[width=0.245\linewidth]{{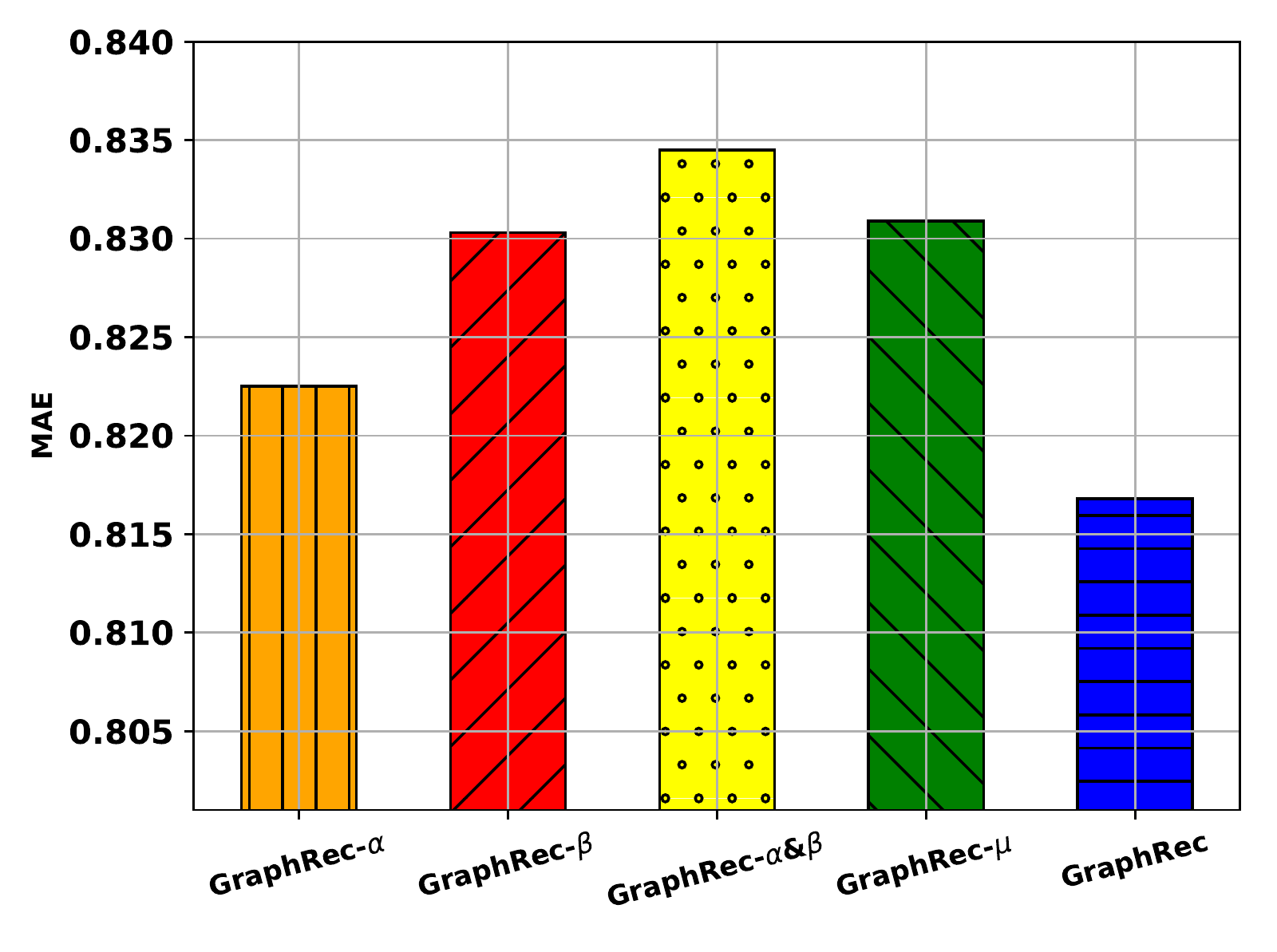}}\label{fig:epinions_att_mae}}}
\caption{Effect of attention mechanisms on Ciao and Epinions datasets.}\label{fig:GraphSR_att}
\end{figure*}

\begin{figure*}[t]
\centering
{\subfigure[Ciao-RMSE]
{\includegraphics[width=0.245\linewidth]{{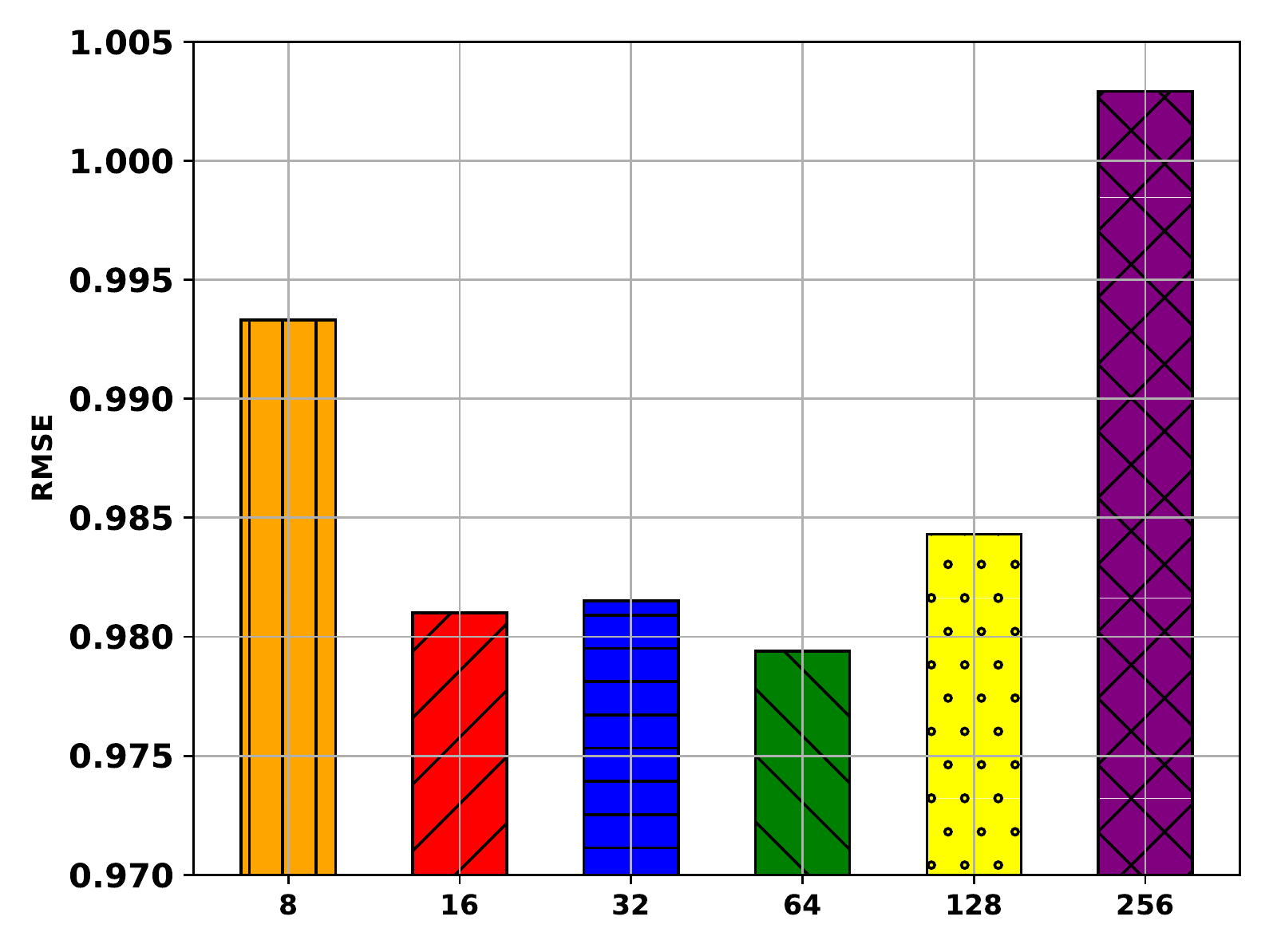}}\label{fig:ciao_emb_rmse}}}
{\subfigure[Ciao-MAE]
{\includegraphics[width=0.245\linewidth]{{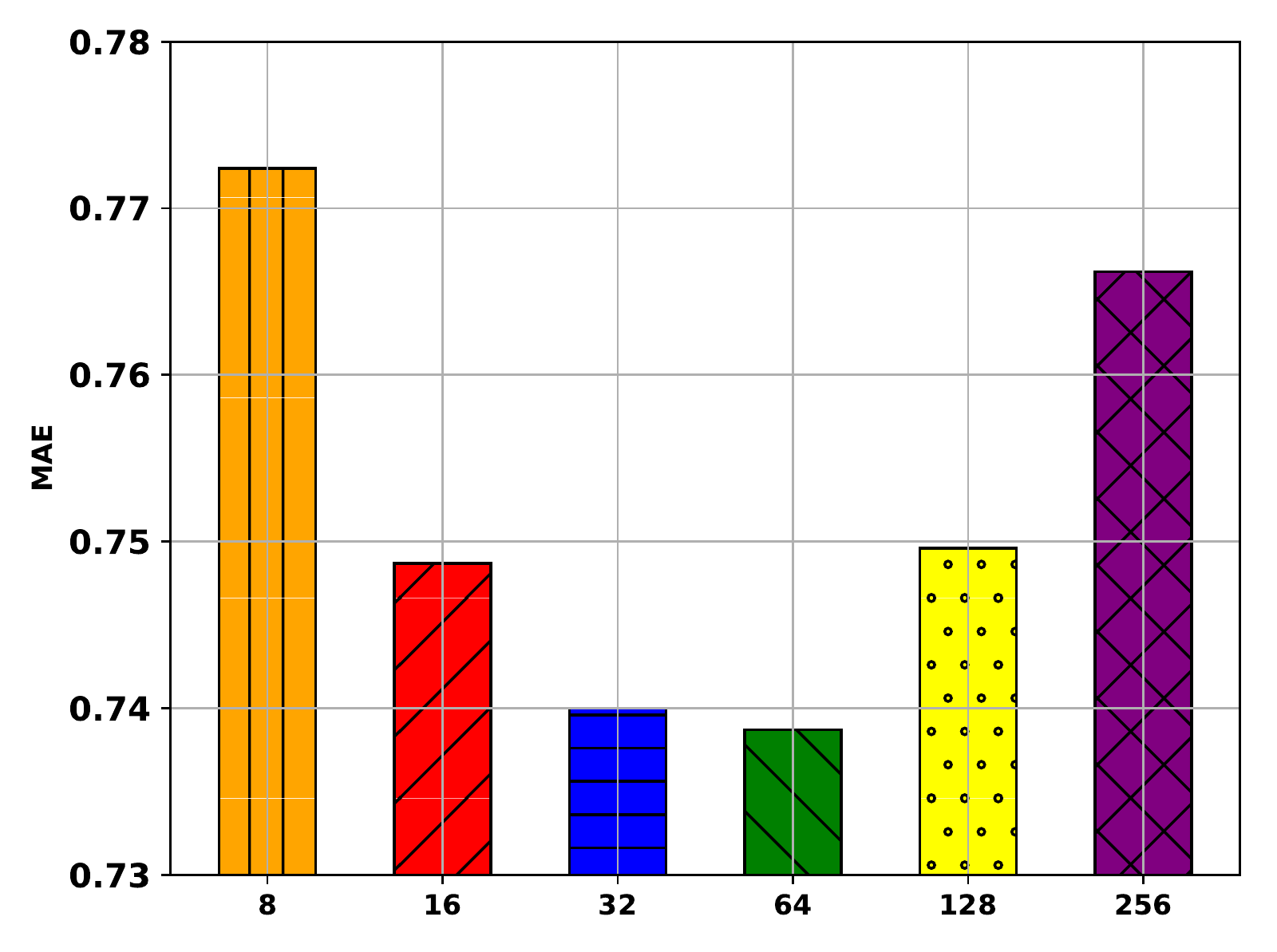}}\label{fig:ciao_emb_mae}}}
{\subfigure[Epinions-RMSE]
{\includegraphics[width=0.245\linewidth]{{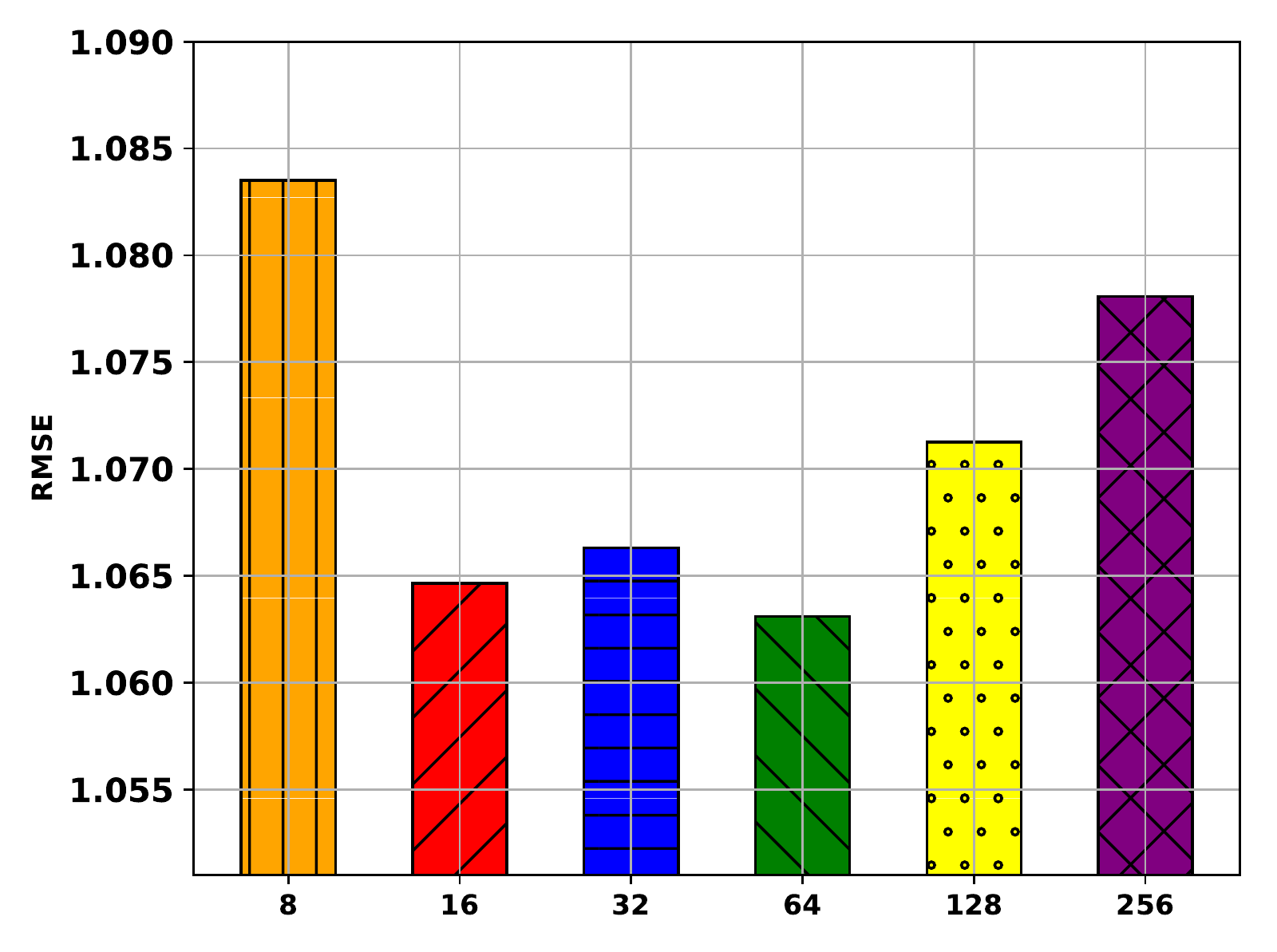}}\label{fig:epinions_emb_rmse}}}
{\subfigure[Epinions-MAE]
{\includegraphics[width=0.245\linewidth]{{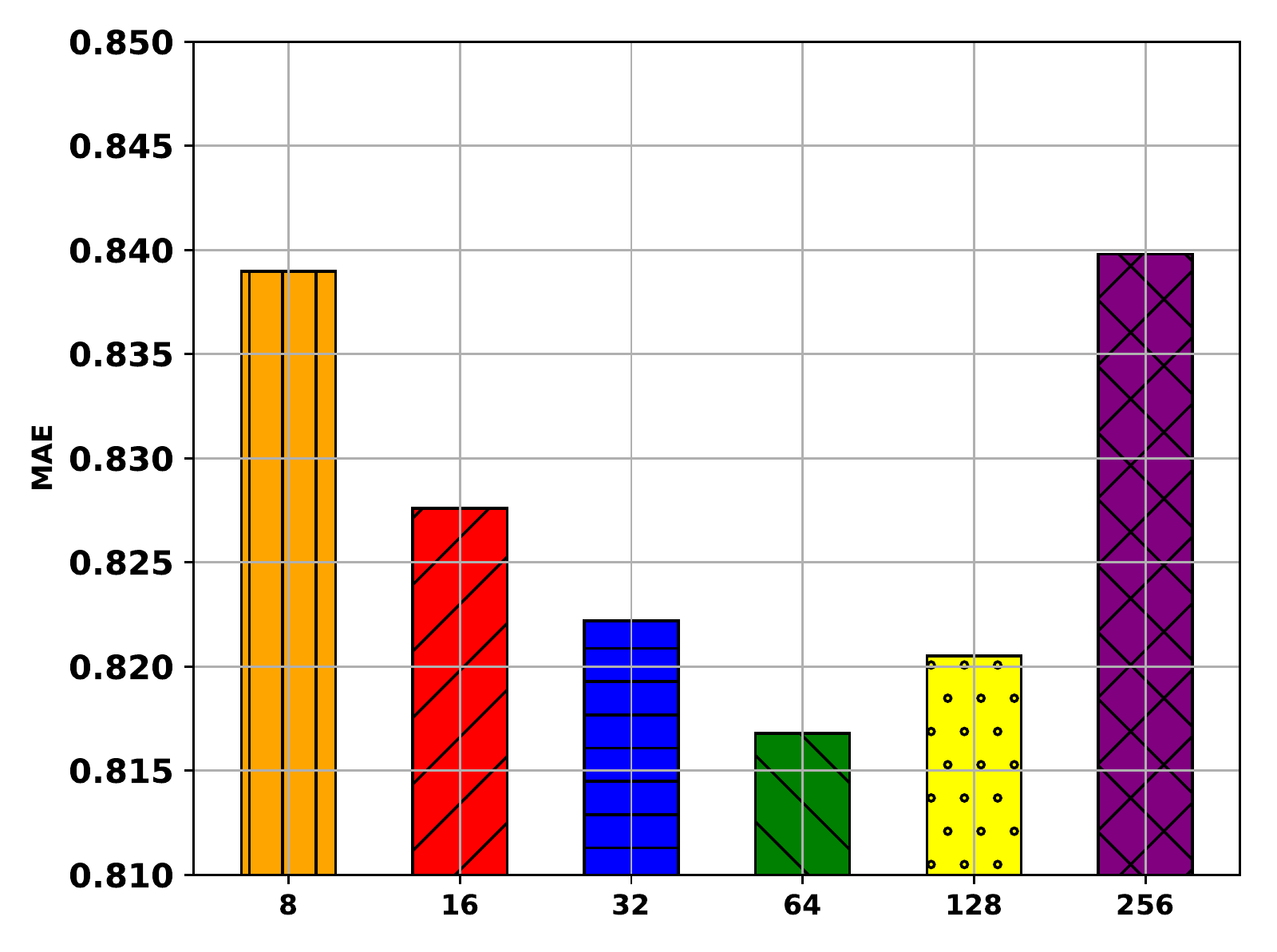}}\label{fig:epinions_emb_mae}}}
\caption{Effect of embedding size  on Ciao and Epinions datasets.}\label{fig:GraphSR_emb}
\end{figure*}

\subsubsection{Effect of Attention Mechanisms}
\label{sec: exp_att}
To get a better understanding of the proposed GraphRec model, we further evaluate the key components of GraphRec - \emph{Attention mechanisms}. There are three different attention mechanisms during aggregation, including item attention $\alpha$, social attention $\beta$, and user attention $\mu$. We compare GraphRec with its four variants: GraphRec-$\alpha$, GraphRec-$\beta$, GraphRec-$\alpha \& \beta$, and GraphRec-$\mu$. These  four variants are defined in the following:

\begin{itemize}
    \item GraphRec-$\alpha$: The item attention $\alpha$ of GraphRec is eliminated during aggregating the opinion-aware interaction representation of items. This variant employs the mean-based aggregation function on item aggregation for modeling item-space user latent factors.
    \item GraphRec-$\beta$: The social attention $\alpha$ is to model users' tie strengths. The social attention $\alpha$ of GraphRec in this variant is eliminated during aggregating user's neighbors. This variant employs the mean-based aggregation function on social aggregation for modeling social-space user latent factors.
    \item GraphRec-$\alpha \& \beta$: This variant eliminates two attention mechanisms (item attention $\alpha$ and social attention $\beta$) on item aggregation and social aggregation for modeling user latent factors. 
    \item GraphRec-$\mu$: The user attention $\mu$ of GraphRec is eliminated during aggregating opinion-aware interaction user representation. This variant employs the mean-based aggregation function on user aggregation for modeling item latent factors.
\end{itemize}

The results of different attention mechanisms on GraphRec are shown in Figure~\ref{fig:GraphSR_att}. From the results, we have the following findings,

\begin{itemize}
  \item Not all interacted items (purchased history) of one user contribute equally to the item-space user latent factor, and not all interacted users (buyers) have the same importance to learning item latent factor. Based on these assumptions, our model considers these difference among users and items by using two different attention mechanisms ($\alpha$ and $\mu$). From the results, we can observe that GraphRec-$\alpha$ and GraphRec-$\mu$ obtain worse performance than GraphRec. These results demonstrate the benefits of the attention mechanisms on item aggregation and user aggregation.
  
  \item As mentioned before, users are likely to share more similar tastes with strong ties than weak ties. The attention mechanism $\beta$ at social aggregation considers heterogeneous strengths of social relations. When the attention mechanism $\beta$ is removed, the performance of GraphRec-$\beta$ is dropped significantly. It justifies our assumption that during social aggregation, different social friends should have different influence for learning social-space user latent factor. It's important to distinguish social relations with heterogeneous strengths.
  
\end{itemize}

To sum up, GraphRec can capture the heterogeneity in aggregation operations of the proposed framework via attention mechanisms, which can boost the recommendation performance.

\subsubsection{Effect of Embedding Size}
In this subsection, to analyze the effect of embedding size  of user embedding $\mathbf{p}$ , item embedding $\mathbf{q}$, and opinion embedding $\mathbf{e}$, on the performance of our model. 

Figure~\ref{fig:GraphSR_emb} presents the performance comparison $w.r.t.$ the length of embedding of our proposed model on Ciao and Epinions datasets. In general, with the increase of the embedding size, the performance first increases and then decreases. When increasing the embedding size from 8 to 64 can improve the performance significantly. However, with the embedding size of 256, GraphRec degrades the performance. It demonstrates that using a large number of the embedding size has powerful representation. Nevertheless, if the length of embedding is too large, the complexity of our model will significantly increase. Therefore, we need to find a proper length of embedding in order to balance the trade-off between the performance and the complexity.

\section{Related work}
\label{sec:relatedwork}

In this section, we briefly review some related work about social recommendation, deep neural network techniques employed for recommendation, and the advanced graph neural networks.

Exploiting social relations for recommendations has attracted significant attention in recent years~\cite{tang2016recommendations, tang2013exploiting,yang2017social}. One common assumption about these models is that a user's preference is similar to or influenced by the people around him/her (nearest neighbours), which can be proven by social correlation theories~\cite{marsden1993network, mcpherson2001birds}. Along with this line, SoRec~\cite{ma2008sorec} proposed a co-factorization method, which shares a common latent user-feature matrix factorized by ratings and by social relations.  TrustMF~\cite{yang2017social} modeled mutual influence between users, and mapped users into two low-dimensional spaces: truster space and trustee space, by factorizing social trust networks.
SoDimRec~\cite{tang2016recommendation} first adopted a community detection algorithm to partition users into several clusters, and then exploited the heterogeneity of social relations and weak dependency connections for recommendation. Comprehensive overviews on social recommender systems can be found in surveys~\citep{tang2013social}.

In recent years, deep neural network models had a great impact on learning effective feature representations in various fields, such as speech recognition~\cite{hinton2012deep}, Computer Vision (CV)~\cite{karimi2018toward} and Natural Language Processing (NLP)~\cite{chen2017survey}. Some recent efforts have applied deep neural networks to recommendation tasks and shown promising results~\cite{zhao2018recommendations}, but most of them used deep neural networks to model audio features of music~\cite{van2013deep}, textual description of items~\cite{wang2015collaborative, Chen2018Neural}, and visual content of images~\cite{ZhaoLP016}. Besides, NeuMF~\cite{He2017NCF} presented a Neural Collaborative Filtering framework to learn the non-linear interactions between users and items.

However, the application of deep neural network in social recommender systems is rare until very recently. In particular, NSCR~\cite{wang2017item} extended the NeuMF~\cite{He2017NCF} model to cross-domain social recommendations, i.e., recommending items of information domains to potential users of social networks, and presented a neural social collaborative ranking recommender system.  However, the limitation is NSCR requires users with one or more social networks accounts (e.g., Facebook, Twitter, Instagram), which limits the data collections and its applications in practice. SMR-MNRL~\cite{zhao2018social} developed social-aware movie recommendation in social media from the viewpoint of learning a multimodal heterogeneous network representation for ranking. They exploited the recurrent neural network and convolutional neural network to learn the representation of movies' textual description and poster image, and adopted a random-walk based learning method into multimodal neural networks. In all these works~\cite{wang2017item}~\cite{zhao2018social}, they addressed the task of cross-domain social recommendations for ranking metric, which is different from traditional social recommender systems.

Most related to our task with neural networks includes DLMF~\cite{deng2017deep} and DeepSoR~\cite{DeepSoR2018}. DLMF~\cite{deng2017deep} used auto-encoder on ratings to learn representation for initializing an existing matrix factorization. A two-phase trust-aware recommendation process is proposed to utilize deep neural networks in matrix factorization's initialization and to synthesize the user's interests and their trust friends' interests together with the impact of community effect based on matrix factorization for recommendations. DeepSoR~\cite{DeepSoR2018}
integrated neural networks for user's social relations into probabilistic matrix factorization. They first represented users using pre-trained node embedding technique, and further exploited k-nearest neighbors to bridge user embedding features and neural network.

More recently, Graph Neural Networks (GNNs) have been proven to be capable of learning on graph structure data~\cite{shuman2013emerging, kipf2017semi, defferrard2016convolutional, derr2018signed, bronstein2017geometric}. In the task of recommender systems, the user-item interaction contains the ratings on items by users, which is a typical graph data. Therefore, GNNs have been proposed to solve the recommendation problem~\cite{monti2017geometric, berg2017graph, Ying:2018:GCN:3219819.3219890}. sRMGCNN~\cite{monti2017geometric} adopted GNNs to extract graph embeddings for users and items, and then combined with recurrent neural network to perform a diffusion process. GCMC~\cite{berg2017graph} proposed a graph auto-encoder framework, which produced latent features of users and items through a form of differentiable message passing on the user-item graph. PinSage~\cite{Ying:2018:GCN:3219819.3219890} proposed a random-walk graph neural network to learn embedding for nodes in web-scale graphs.  Despite the compelling success achieved by previous work, little attention has been paid to social recommendation with GNNs. In this paper, we propose a graph neural network for social recommendation to fill this gap.

\section{Conclusion and Future work}
\label{sec:conclusion}

We have presented a Graph Network model (GraphRec) to model social recommendation for rating prediction. Particularly, we provide a principled approach to jointly capture interactions and opinions in the user-item graph. Our experiments reveal that the opinion information plays a crucial role in the improvement of our model performance. In addition, our GraphRec can differentiate the ties strengths by considering heterogeneous strengths of social relations.
Experimental results on two real-world datasets show that GraphRec can outperform state-of-the-art baselines.

Currently we only incorporate the social graph into recommendation, while many real-world industries are associated rich other side information on users as well as items. For example, users and items are associated with rich attributes. Therefore, exploring graph neural networks for recommendation with attributes would be an interesting future direction. Beyond that, now we consider both rating and social information static. However, rating and social information are naturally dynamic. Hence, we will consider building dynamic graph neural networks for social recommendations with dynamic.

\begin{acks}

The work described in this paper has been supported, in part, by a general research fund from the Hong Kong Research Grants Council (project PolyU 1121417/17E), and an internal research grant from the Hong Kong Polytechnic University (project 1.9B0V). Yao Ma and Jiliang Tang are supported by the National Science Foundation (NSF) under grant numbers IIS-1714741, IIS-1715940 and CNS-1815636, and a grant from Criteo Faculty Research Award.

\end{acks}

\bibliographystyle{ACM-Reference-Format}
\balance
\bibliography{references/references}
\end{document}